\documentclass[a4paper,fleqn,usenatbib]{mnras}
\usepackage[T1]{fontenc}
\usepackage{ae,aecompl}

\usepackage{graphicx}	% Including figure files
\usepackage{amsmath}	% Advanced maths commands
\usepackage{amssymb}	% Extra maths symbols

\usepackage{grffile}
\usepackage{braket}
\title[Minimum variance estimation]{Minimum variance estimation of galaxy power spectrum in redshift space}

\author[M. Shiraishi et al.]{Maresuke Shiraishi,$^{1}$\thanks{E-mail: shiraishi-m@t.kagawa-nct.ac.jp}
Teppei Okumura,$^{2,3}$
Naonori S. Sugiyama$^{4}$,
Kazuyuki Akitsu$^{3}$
\\
$^1$Department of General Education, National Institute of Technology, Kagawa College, \\
355 Chokushi-cho, Takamatsu, Kagawa 761-8058, Japan \\
$^2$Academia Sinica Institute of Astronomy and Astrophysics (ASIAA), No. 1, Section 4, Roosevelt Road, Taipei 10617, Taiwan \\
$^3$Kavli Institute for the Physics and Mathematics of the Universe (WPI), UTIAS, The University of Tokyo, \\
Kashiwa, Chiba 277-8583, Japan \\
$^4$National Astronomical Observatory of Japan, Mitaka, Tokyo 181-8588, Japan
}

\date{Accepted 2020 July 17. Received 2020 July 17; in original form 2020 May 9}

\begin{document}
\label{firstpage}
\pagerange{\pageref{firstpage}--\pageref{lastpage}}
\maketitle

\begin{abstract}

  We study an efficient way to enhance the measurability of the galaxy density and/or velocity power spectrum in redshift space. It is based on the angular decomposition with the Tripolar spherical harmonic (TripoSH) basis and applicable even to galaxy distributions in wide-angle galaxy surveys. While nontrivial multipole-mode mixings are inevitable in the covariance of the Legendre decomposition coefficient commonly used in the small-angle power spectrum analysis, our analytic computation of the covariance of the TripoSH decomposition coefficient shows that such mixings are absent by virtue of high separability of the TripoSH basis, yielding the minimum variance. Via the simple signal-to-noise ratio assessment, we confirm that the detectability improvement by the TripoSH decomposition approach becomes more significant at higher multipole modes, and, e.g., the hexadecapole of the density power spectrum has two orders of magnitude improvement. The TripoSH decomposition approach is expected to be applied to not only currently available survey data but also forthcoming wide-angle one, and to bring about something new or much more accurate cosmological information.
  
\end{abstract}


%%%%%%%%%%%%%%%%%%%%%%%%%%%%%%%%%%%%%%%%%%%%%%%%%%

%%%%%%%%%%%%%%%%% BODY OF PAPER %%%%%%%%%%%%%%%%%%

%%%%%%%%%%%%%%%%%%%%%%%%%%%%%%%%%%%%%%%%%%%%%%%%%%%%%%%%%%%%%%%%%%%%%%%%%%%%%%
\section{Introduction}
%%%%%%%%%%%%%%%%%%%%%%%%%%%%%%%%%%%%%%%%%%%%%%%%%%%%%%%%%%%%%%%%%%%%%%%%%%%%%%

The power spectrum, i.e., the Fourier transform of the two-point correlation function, of redshift-space galaxy density and velocity fields is an indispensable statistic for precise understanding of the Universe \citep{Peebles:1980,Strauss:1995,Hamilton:1997zq}. Thanks to recent precise measurements, various cosmological parameters have been tightly constrained, and the nature of gravity or the particle content of the Universe has been severely tested \citep[see e.g.,][for review]{Weinberg:2012es}.

In most previous data analysis mentioned above, a traditional Legendre-decomposition-based estimator \citep{Yamamoto:2005dz} has been employed. The observed power spectrum is, originally, characterized by two independent angles: $\hat{k} \cdot \hat{s}_1$ and $\hat{k} \cdot \hat{s}_2$, where $\hat{s}_1$ and $\hat{s}_2$ denote the line-of-sight (LOS) directions toward two fields, due to the redshift-space distortion (RSD) \citep{Hamilton:1997zq}. In this estimator, however, these angles are identified with each other as $\hat{k} \cdot \hat{s}_1 = \hat{k} \cdot \hat{s}_2 \equiv \mu$; thus, the angular dependence of the power spectrum is decomposed using the Legendre polynomials ${\cal L}_\ell(\mu)$. This approximation, dubbed the plane-parallel (PP) approximation, makes the data analysis pipeline much simpler. At the same time, however, this gives rise to nontrivial couplings between different $\ell$ modes in the covariance \citep[see e.g.,][]{Taruya:2010mx}, obstructing the minimization of the estimator variance. There are a few studies that performed the cosmological analysis of observed galaxy clustering without relying on the PP approximation in configuration space \citep[e.g.,][]{Matsubara:2000a,Pope:2004,Okumura:2008}.

In this Letter, we propose a way to achieve the minimum variance. It is done by the angular decomposition using the tripolar spherical harmonic (TripoSH) basis $\{ Y_{\ell}(\hat{k}) \otimes \{ Y_{\ell_1}(\hat{s}_1) \otimes Y_{\ell_2}(\hat{s}_2) \}_{\ell} \}_{00}$~\citep{Varshalovich:1988ye}. Its crucial difference from the Legendre polynomial basis is that there are two additional multipoles $\ell_1$ and $\ell_2$ that can decompose the dependences on $\hat{k} \cdot \hat{s}_1$ and $\hat{k} \cdot \hat{s}_2$ separably. The remaining multipole $\ell$ represents the total angular momentum of $\ell_1$ and $\ell_2$, and is identified with the Legendre multipole under the PP approximation [see Eq.~\eqref{eq:calP_2_P} below]. There are already numerous studies on the responses of the TripoSH decomposition coefficients on various theoretical power spectrum templates \citep[e.g.,][]{Szalay:1997cc,Szapudi:2004gh,Papai:2008bd,Bertacca:2012tp,Yoo:2013zga,Shiraishi:2016wec}, while computing their covariances and signal-to-noise ratios (SNRs) is, for the first time, done in this Letter. Note that our decomposition approach does not rely on the PP approximation and is therefore applicable even to the wide-angle galaxy distribution.%
\footnote{See e.g., \citet{Castorina:2017inr,Beutler:2018vpe,Castorina:2019hyr} for studies on other decomposition approaches to the wide-angle effect.}
We then find that, owing to high separability of the TripoSH basis, each $\ell$ mode does not tangle even at the covariance level and the minimum variance is consequently realized. We also compare the SNR of the TripoSH coefficient with that of the Legendre one for each $\ell$ mode, and test how much the detectability is improved by use of the TripoSH coefficient. We show that the gain of the SNR is found to be more significant for higher $\ell$ modes, and for example, the two orders of magnitude improvement is found for the $\ell=4$ (hexadecapole) mode.

%%%%%%%%%%%%%%%%%%%%%%%%%%%%%%%%%%%%%%%%%%%%%%%%%%%%%%%%%%%%%%%%%%%%%%%%%%%%%%
\section{Preliminaries} \label{sec:linear} 
%%%%%%%%%%%%%%%%%%%%%%%%%%%%%%%%%%%%%%%%%%%%%%%%%%%%%%%%%%%%%%%%%%%%%%%%%%%%%%

All the analyses in this Letter are done on the basis of the linear theory because we are primarily interested in large-scale galaxy clustering; therefore, the galaxy number density fluctuation, $\delta({\bf s}) \equiv n({\bf s}) / \bar{n}(s) - 1 $, and the LOS peculiar velocity field, $u({\bf s}) \equiv {\bf v}({\bf s}) \cdot \hat{s}$, in redshift space are expressed as \citep{Hamilton:1997zq,Yoo:2013zga}
\begin{eqnarray}
  X ({\bf s}) = \int \frac{d^3 k}{(2\pi)^3} e^{i {\bf k} \cdot {\bf s}} F^X({\bf k}, \hat{s}) ,
\end{eqnarray}
where $X = \{ \delta, u \}$ and 
\begin{eqnarray}
  \begin{split}
    F^\delta({\bf k}, \hat{s}) &\equiv
\left[ b  - 
i \frac{\alpha(s)}{ks} (\hat{k} \cdot \hat{s}) f  
+ (\hat{k} \cdot \hat{s})^2 f  \right] \delta_m({\bf k}) 
, \\
%---
F^u({\bf k}, \hat{s}) &\equiv i \frac{aH}{k} (\hat{k} \cdot \hat{s}) f \delta_m ({\bf k}) ,
\end{split}
\end{eqnarray}
with $b$ the linear bias parameter, $f$ the linear growth rate, $\delta_m$ the matter density perturbation in real space, $a$ the scale factor and $H$ the Hubble parameter. The selection function $\alpha$ is defined in terms of the mean number density $\bar{n}(s)$ of the galaxy sample as $\alpha(s) \equiv d \ln \bar{n}(s) / d \ln s + 2$. Here and hereinafter, we omit time, redshift, or the radial distance in the arguments of variables unless the parameter dependence is nontrivial. For convenience, let us rewrite these by use of the Legendre polynomials as 
\begin{eqnarray}
  F^X({\bf k}, \hat{s}) = \sum_{j} c_j^X(k) {\cal L}_j(\hat{k} \cdot \hat{s}) \delta_m({\bf k}) ,
  \end{eqnarray}
with 
\begin{eqnarray}
  \begin{split}
    & c_0^\delta = b + \frac{1}{3} f   , \ \
  c_1^\delta = - i \frac{\alpha}{k s} f , \ \
  c_2^\delta = \frac{2}{3} f , \ \
  c_{j \geq 3}^{\delta} = 0 , \\
  %---
  & c_1^u = i \frac{aH}{k} f, \ \
  c_0^u = c_{j \geq 2}^u = 0 .
  \end{split} \label{eq:delta_u_coeff}
\end{eqnarray}
Note that $F^X({\bf k}, \hat{s})$ is not equal to the Fourier counterpart of $X({\bf s})$ because there still remains the position dependence.

Throughout this Letter, we assume that the real-space matter power spectrum takes a statistically homogeneous and isotropic form:
\begin{eqnarray}
  \Braket{ \delta_m({\bf k}_1) \delta_m({\bf k}_2) } = (2\pi)^3 \delta^{(3)}({\bf k}_1 + {\bf k}_2) P_m(k_1). 
\end{eqnarray}
The redshift-space two-point correlation function then takes the form:
\begin{eqnarray}
  \begin{split}
  \xi^{X_1 X_2}({\bf s}_{12}, \hat{s}_1, \hat{s}_2)
  &\equiv \Braket{X_1({\bf s}_1) X_2({\bf s}_2)} \\ 
  &= \int \frac{d^3 k}{(2\pi)^3} e^{i {\bf k} \cdot {\bf s}_{12}} P^{X_1 X_2}({\bf k}, \hat{s}_1, \hat{s}_2) , 
  \end{split} \label{eq:xi_homo}
\end{eqnarray}
where ${\bf s}_{12} \equiv {\bf s}_1 - {\bf s}_2$ and
\begin{eqnarray}
  \begin{split}
    P^{X_1 X_2}({\bf k}, \hat{s}_1, \hat{s}_2)
    &\equiv  \sum_{j_1 j_2} c_{j_1}^{X_1}(k) (-1)^{j_2}c_{j_2}^{X_2} (k)  \\ 
    &\quad \times {\cal L}_{j_1}(\hat{k} \cdot \hat{s}_1) 
    {\cal L}_{j_2}(\hat{k} \cdot \hat{s}_2)
    P_m(k).
  \end{split}
  \label{eq:P_homo}
\end{eqnarray}
Again, this $P^{X_1 X_2}$ differs from the Fourier counterpart of $\xi^{X_1 X_2}$ and they should not be confused with each other.

In this Letter, we also do not take unequal time correlators into account although these are also nonzero and should enhance the SNR; therefore, the radial distances appearing in $c_1^\delta$ are fixed to be $s_1 \simeq s_2$.

%%%%%%%%%%%%%%%%%%%%%%%%%%%%%%%%%%%%%%%%%%%%%%%%%%%%%%%%%%%%%%%%%%%%%%%%%%%%%%
\section{TripoSH decomposition} \label{sec:TriposH}
%%%%%%%%%%%%%%%%%%%%%%%%%%%%%%%%%%%%%%%%%%%%%%%%%%%%%%%%%%%%%%%%%%%%%%%%%%%%%%

Let us introduce the TripoSH basis with zero total angular momentum using the Wigner 3$j$ symbol \citep{Varshalovich:1988ye}:
\begin{eqnarray}
  \begin{split}
    {\cal X}_{\ell \ell_1\ell_2}(\hat{s}_{12},\hat{s}_1,\hat{s}_2)
    &\equiv \{ Y_{\ell}(\hat{s}_{12}) \otimes \{ Y_{\ell_1}(\hat{s}_1) \otimes Y_{\ell_2}(\hat{s}_2) \}_{\ell} \}_{00} \\
    %@@@
    &= \sum_{m m_1 m_2} 
    (-1)^{\ell_1 + \ell_2 + \ell}
    \left( \begin{matrix}
      \ell_1 & \ell_2 & \ell \\
      m_1 & m_2 & m
    \end{matrix}  \right) \\ 
    &\quad \times Y_{\ell m}(\hat{s}_{12}) Y_{\ell_1 m_1}(\hat{s}_1) Y_{\ell_2 m_2}(\hat{s}_2).
  \end{split}
  \label{eq:Xbasis_def}
\end{eqnarray}
The TripoSH decomposition is done according to
\begin{eqnarray}
  \xi^{X_1 X_2}({\bf s}_{12}, \hat{s}_1, \hat{s}_2)
  = \sum_{\ell\ell_1\ell_2} \Xi_{\ell\ell_1\ell_2}^{X_1 X_2}(s_{12}) 
  {\cal X}_{\ell\ell_1\ell_2}(\hat{s}_{12},\hat{s}_1,\hat{s}_2).
\end{eqnarray}
In this process, the dependences on $\hat{s}_1$ and $\hat{s}_2$ are characterized by $\ell_1$ and $\ell_2$, respectively, and $\ell$ denotes the total angular momentum of $\ell_1$ and $\ell_2$. The orthonormality of ${\cal X}_{\ell \ell_1\ell_2}(\hat{s}_{12},\hat{s}_1,\hat{s}_2)$ yields an inverse formula:
\begin{eqnarray}
  \begin{split}
\Xi_{\ell \ell_1 \ell_2}^{X_1 X_2}(s_{12}) 
  &=  \int d^2 \hat{s}_{12}  \int d^2 \hat{s}_1  \int d^2 \hat{s}_2
\xi^{X_1 X_2}({\bf s}_{12}, \hat{s}_1, \hat{s}_2) \\
&\quad \times {\cal X}_{\ell \ell_1\ell_2}^{*}(\hat{s}_{12},\hat{s}_1,\hat{s}_2).
  \end{split}
  \label{eq:Xi_def}
\end{eqnarray}
This is connected to the decomposition coefficients, defined by
\begin{eqnarray}
  \begin{split}
    \Pi_{\ell \ell_1 \ell_2}^{X_1 X_2}(k) 
    &\equiv  \int d^2 \hat{k}  \int d^2 \hat{s}_1  \int d^2 \hat{s}_2
    P^{X_1 X_2}({\bf k}, \hat{s}_1, \hat{s}_2) \\
    &\quad \times {\cal X}_{\ell \ell_1\ell_2}^{*}(\hat{k},\hat{s}_1,\hat{s}_2) ,
  \end{split}
  \label{eq:Pi_def}
\end{eqnarray}
via the Hankel transformation:
\begin{eqnarray}
  \Xi_{\ell\ell_1\ell_2}^{X_1 X_2}(s_{12}) 
  = i^{\ell} \int_0^\infty \frac{k^2 dk}{2\pi^2} j_{\ell}(k s_{12})
  \Pi_{\ell \ell_1 \ell_2}^{X_1 X_2}(k) ,  \label{eq:hankel}
\end{eqnarray}
with $j_{\ell}(x)$ the spherical Bessel functions.

In the following, for convenience, the TripoSH coefficients, obtained from Eq.~\eqref{eq:Pi_def}, are utilized via an additional transformation:
\begin{eqnarray}
  {\cal P}_{\ell \ell_1 \ell_2}^{X_1 X_2}(k) 
  \equiv \frac{h_{\ell_1 \ell_2 \ell}}{4\pi}
  \Pi_{\ell \ell_1 \ell_2}^{X_1 X_2}(k) , \label{eq:calP_def}
\end{eqnarray}
where
\begin{eqnarray}
h_{l_1 l_2 l_3} \equiv \sqrt{\frac{(2 l_1 + 1)(2 l_2 + 1)(2 l_3 + 1)}{4 \pi}} \left(\begin{matrix}
  l_1 & l_2 & l_3 \\
   0 & 0 & 0 
\end{matrix}\right).
\end{eqnarray}
Taking the PP limit ($\hat{s}_1 = \hat{s}_2 = \hat{s}$), these reduced coefficients are related to the conventional Legendre decomposition ones through a simple summation:
\begin{eqnarray}
  P_{\ell}^{X_1 X_2} = \sum_{\ell_1 \ell_2} {\cal P}_{\ell \ell_1 \ell_2}^{X_1 X_2}; \label{eq:calP_2_P}
\end{eqnarray}
therefore, are useful especially for the comparison with the PP-limit results. Note that this is derived employing an identity ${\cal X}_{\ell \ell_1\ell_2}(\hat{k},\hat{s},\hat{s}) = h_{\ell_1 \ell_2 \ell} {\cal L}_\ell(\hat{k} \cdot \hat{s}) / (4\pi)$.

The above decomposition scheme follows the same format as the previous studies on the density field \citep[e.g.,][]{Szapudi:2004gh,Shiraishi:2016wec}, and is newly extended to the analysis of the velocity field here. Performing the angular integrals in Eq.~\eqref{eq:Pi_def}, the power spectrum signal is decomposed into the TripoSH coefficients. In our case, as the power spectrum takes the simple form~\eqref{eq:P_homo}, these are done analytically as
\begin{eqnarray}
  \begin{split}
  & \int d^2 \hat{s}_1  \int d^2 \hat{s}_2
   {\cal X}_{\ell \ell_1\ell_2 }^{ *}(\hat{k},\hat{s}_1,\hat{s}_2)
   {\cal L}_{j_1}(\hat{k} \cdot \hat{s}_1)
   {\cal L}_{j_2}(\hat{k} \cdot \hat{s}_2) \\ 
   & \quad = \frac{4\pi h_{\ell_1 \ell_2 \ell}}{(2\ell_1 + 1)(2\ell_2 + 1)}  
   \delta_{\ell_1, j_1} \delta_{\ell_2, j_2}.  
  \end{split}
  \label{eq:int_XLL}
\end{eqnarray}
We then finally obtain
\begin{eqnarray}
  {\cal P}_{\ell \ell_1 \ell_2}^{X_1 X_2}(k)
  = \frac{4\pi (-1)^{\ell_2} h_{\ell_1 \ell_2 \ell}^2 }{(2\ell_1 + 1)(2\ell_2 + 1)} 
  c_{\ell_1}^{X_1}(k) c_{\ell_2}^{X_2}(k) P_m(k) .
\end{eqnarray}
Due to the selection rule of $h_{\ell_1 \ell_2 \ell}$, i.e., $ |\ell_1 - \ell_2| \leq \ell \leq \ell_1 + \ell_2$ and $\ell_1 + \ell_2 + \ell = {\rm even}$, and a nonvanishing condition of $c_\ell^X$ [see Eq.~\eqref{eq:delta_u_coeff}], only 14~${\cal P}_{\ell \ell_1 \ell_2}^{\delta \delta}$, 5~${\cal P}_{\ell \ell_1 \ell_2}^{\delta u}$ and 2~${\cal P}_{\ell \ell_1 \ell_2}^{uu}$ listed below can take nonzero values:
\begin{eqnarray}
  \begin{cases}
    {\cal P}_{000}^{\delta \delta} , \ 
    {\cal P}_{011}^{\delta \delta} , \  
    {\cal P}_{022}^{\delta \delta} , \\
    %@@@
    {\cal P}_{101}^{\delta \delta} , \ 
    {\cal P}_{110}^{\delta \delta} , \ 
    {\cal P}_{112}^{\delta \delta} , \
    {\cal P}_{121}^{\delta \delta} , \\ 
    %@@@
    {\cal P}_{202}^{\delta \delta} , \ 
    {\cal P}_{211}^{\delta \delta} , \  
    {\cal P}_{220}^{\delta \delta} , \  
    {\cal P}_{222}^{\delta \delta} , \\
    %@@@
    {\cal P}_{312}^{\delta \delta} , \ 
    {\cal P}_{321}^{\delta \delta} , \\ 
    %@@@
    {\cal P}_{422}^{\delta \delta} ,
  \end{cases} 
    %---
  \begin{cases}
    {\cal P}_{011}^{\delta u} , \\
    %@@@
    {\cal P}_{101}^{\delta u} , \
    {\cal P}_{121}^{\delta u} , \\
    %@@@
    {\cal P}_{211}^{\delta u} , \\
    %@@@
    {\cal P}_{321}^{\delta u} ,
  \end{cases}
    %---
  \begin{cases} 
    {\cal P}_{011}^{u u} , \\
    %@@@
    {\cal P}_{211}^{u u} .
  \end{cases}
  \label{eq:calP_list}
\end{eqnarray}
Note that, since the transformation \eqref{eq:calP_def} is done without loss of generality, $\Pi_{\ell \ell_1 \ell_2}^{X_1 X_2}$ has the equivalent information and thus takes nonzero values only at the same multipole configurations.

%%%%%%%%%%%%%%%%%%%%%%%%%%%%%%%%%%%%%%%%%%%%%%%%%%%%%%%%%%%%%%%%%%%%%%%%%%%%%%
\section{TripoSH covariance} \label{sec:cov}
%%%%%%%%%%%%%%%%%%%%%%%%%%%%%%%%%%%%%%%%%%%%%%%%%%%%%%%%%%%%%%%%%%%%%%%%%%%%%%

 It is expected from Eq.~\eqref{eq:xi_homo} that $\xi^{X_1 X_2}$ extracted from given data takes the form:
\begin{eqnarray}
\hat{\xi}^{X_1 X_2}({\bf s}_{12}, \hat{s}_1, \hat{s}_2) 
  = \int \frac{d^3 k}{(2\pi)^3} e^{i {\bf k} \cdot {\bf s}_{12}} \hat{P}^{X_1 X_2}({\bf k}, \hat{s}_1, \hat{s}_2) ,
  \label{eq:xi_P}
\end{eqnarray}
where $\hat{P}^{X_1 X_2}({\bf k}, \hat{s}_1, \hat{s}_2) \equiv F^{X_1}({\bf k}, \hat{s}_1) F^{X_2}(-{\bf k}, \hat{s}_2) / V - P_{\rm noise}^{X_1 X_2} $ with $V$ the survey volume and $P_{\rm noise}^{X_1 X_2}$ the contribution of the shot noise. Here and hereinafter, $\xi^{X_1 X_2}$, $P^{X_1 X_2}$ and their decomposition coefficients with hat denote quantities calculated from a single realization, and they should not be confused with the unit vector. Supposing Gaussianity of $F^X$, the covariance of $\hat{P}^{X_1 X_2}$ is computed as 
\begin{eqnarray}
  \begin{split}
    & \Braket{ \hat{P}^{X_1 X_2}({\bf k}, \hat{s}_1, \hat{s}_2)) \hat{P}^{\tilde{X}_1 \tilde{X}_2}(\tilde{\bf k}, \hat{\tilde{s}}_1, \hat{\tilde{s}}_2)) }_c
    = 
    \\
    & 4\pi \frac{\delta_{k,\tilde{k}}}{N_k} 
  \left[
     \delta^{(2)}(\hat{k} + \hat{\tilde{k}})
    P_{\rm tot}^{X_1 \tilde{X}_1} ({\bf k}, \hat{s}_1, \hat{\tilde{s}}_1 )
    P_{\rm tot}^{X_2 \tilde{X}_2} (-{\bf k}, \hat{s}_2, \hat{\tilde{s}}_2 ) \right.  \\
   & \left. \qquad
    + \delta^{(2)}(\hat{k} - \hat{\tilde{k}}) P_{\rm tot}^{X_1 \tilde{X}_2}({\bf k}, \hat{s}_1, \hat{\tilde{s}}_2 )
    P_{\rm tot}^{X_2 \tilde{X}_1}(-{\bf k}, \hat{s}_2, \hat{\tilde{s}}_1 )  \right],
  \end{split}
  \label{eq:P_covmat}
\end{eqnarray}
where $N_k = V k^2 dk / (2\pi^2)$ and $P_{\rm tot}^{X_1 X_2} \equiv P^{X_1 X_2} + P_{\rm noise}^{X_1 X_2}$. It is convenient to rewrite $P_{\rm tot}^{X_1 X_2}$ into
\begin{equation}
  P_{\rm tot}^{X_1 X_2} ({\bf k}, \hat{s}_1, \hat{s}_2 )
  = \sum_{j_1 j_2} (-1)^{j_2} p_{j_1 j_2}^{X_1 X_2}(k)  {\cal L}_{j_1}(\hat{k} \cdot \hat{s}_1) {\cal L}_{j_2}(\hat{k} \cdot \hat{s}_2) ,
\end{equation}
where
\begin{eqnarray}
  p_{j_1 j_2}^{X_1 X_2}(k) = c_{j_1}^{X_1}(k) c_{j_2}^{X_2} (k) P_m(k) 
  + P_{\rm noise}^{X_1 X_2}  \delta_{j_1, 0} \delta_{j_2, 0} .
\end{eqnarray}

The covariance of $\hat{\cal P}_{\ell \ell_1 \ell_2 }^{X_1 X_2}$ is obtained via the double TripoSH decomposition of Eq.~\eqref{eq:P_covmat}. Again, the resulting angular integrals can be simplified so much owing to the analytic formula~\eqref{eq:int_XLL}. The bottom-line form reads
\begin{eqnarray}
  \Braket{\hat{\cal P}_{\ell \ell_1 \ell_2}^{X_1 X_2}(k) \hat{\cal P}_{\tilde{\ell} \tilde{\ell}_1 \tilde{\ell}_2 }^{\tilde{X}_1 \tilde{X}_2 *}(\tilde{k})  }_c  
  = \frac{\delta_{k,\tilde{k}}}{N_k}
  \Upsilon_{\ell \ell_1\ell_2 ; \tilde{\ell} \tilde{\ell}_1 \tilde{\ell}_2 }^{X_1 X_2 ; \tilde{X}_1 \tilde{X}_2}(k) ,
  \label{eq:cov}
\end{eqnarray}
where
\begin{eqnarray}
  \begin{split}
    & \Upsilon_{\ell \ell_1\ell_2 ; \tilde{\ell} \tilde{\ell}_1 \tilde{\ell}_2 }^{X_1 X_2 ; \tilde{X}_1 \tilde{X}_2}(k)
  \equiv \frac{(4\pi)^2 (-1)^{\ell_2 + \tilde{\ell}_1} h_{\ell_1 \ell_2 \ell}^2 h_{\tilde{\ell}_1 \tilde{\ell}_2 \tilde{\ell}}^2 }{(2\ell_1 + 1)(2\ell_2 + 1)(2 \tilde{\ell}_1 + 1)(2 \tilde{\ell}_2 + 1)}  \\ 
  &\qquad\quad \times 
  \left[ p_{\ell_1 \tilde{\ell}_1}^{X_1 \tilde{X}_1}(k)
    p_{\ell_2 \tilde{\ell}_2}^{X_2 \tilde{X}_2}(k)
    + 
    p_{\ell_1 \tilde{\ell}_2}^{X_1 \tilde{X}_2}(k)
    p_{\ell_2 \tilde{\ell}_1}^{X_2 \tilde{X}_1}(k)
    \right] .
  \end{split}
\end{eqnarray}
The covariance of $\hat{\Xi}_{\ell \ell_1 \ell_2}^{X_1 X_2}$ is obtained by following the transformations \eqref{eq:hankel} and \eqref{eq:calP_def}.
One very interesting finding from this expression is that, at some specific multipole configurations, e.g., where none or only one of $\ell_1$, $\ell_2$, $\tilde{\ell}_1$ and $\tilde{\ell}_2$ takes 0, the covariance is free from the shot noise term. In such a situation, the covariance is minimized as
\begin{eqnarray}
  \Upsilon_{\ell \ell_1\ell_2 ; \tilde{\ell} \tilde{\ell}_1 \tilde{\ell}_2 }^{X_1 X_2 ; \tilde{X}_1 \tilde{X}_2}(k)
  = 2 {\cal P}_{\ell \ell_1 \ell_2}^{X_1 X_2}(k) {\cal P}_{\tilde{\ell} \tilde{\ell}_1 \tilde{\ell}_2}^{\tilde{X}_1 \tilde{X}_2 *}(k). \label{eq:Upsilon_CVL}
\end{eqnarray}

We stress that this is not the case if one uses the Legendre decomposition, which is adopted by most of preceding studies,
\begin{eqnarray}
    \hat{P}_\ell^{X_1 X_2}(k) = \frac{2\ell + 1}{2} \int_{-1}^1 d(\hat{k} \cdot \hat{s}) \hat{P}^{X_1 X_2}({\bf k}, \hat{s}, \hat{s})  {\cal L}_\ell(\hat{k} \cdot \hat{s}) .
\end{eqnarray}
The identification between $\hat{k} \cdot \hat{s}_1$ and $\hat{k} \cdot \hat{s}_2$ causes nontrivial mode mixings between different $\ell$ modes at the covariance level. Let us focus on $ \hat{P}_\ell^{uu}$ for example. The Legendre decomposition of $P^{uu}$ under the PP approximation yields two different nonvanishing multipoles: $P_0^{uu}$ and $P_2^{uu}$, while the covariance of $\hat{P}_2^{uu}$ contains not only $P_2^{uu}$ but also $P_{0}^{u u} + P_{\rm noise}^{uu}$ as
\begin{eqnarray}
  \begin{split}
    & \Braket{\hat{P}_{2}^{uu}(k) \hat{P}_{2}^{uu *}(\tilde{k})}_c 
  = \frac{\delta_{k,\tilde{k}}}{N_k} \left[ 10 \left( P_{0}^{u u}(k) + P_{\rm noise}^{uu} \right)^2 \right.  \\
    & \left. \qquad + \frac{40}{7} \left( P_{0}^{u u}(k) + P_{\rm noise}^{uu} \right) P_2^{u u}(k)
    + \frac{30}{7} \left( P_2^{u u}(k) \right)^2  \right].
  \end{split}
\end{eqnarray}
On the other hand, as is evident from Eq.~\eqref{eq:Upsilon_CVL}, the covariance of $\hat{\cal P}_{2 1 1}^{uu}$ is given by ${\cal P}_{2 1 1}^{uu}$ alone.
The similar different-mode contamination also occurs in the covariance of $\hat{P}_\ell^{\delta u}$ and $\hat{P}_\ell^{\delta \delta}$ \citep[see e.g.,][for practical expressions]{Taruya:2010mx}. This causes the loss of detectability.

%%%%%%%%%%%%%%%%%%%%%%%%%%%%%%%%%%%%%%%%%%%%%%%%%%%%%%%%%%%%%%%%%%%%%%%%%%%%%%
\section{Efficiency} \label{sec:SN}
%%%%%%%%%%%%%%%%%%%%%%%%%%%%%%%%%%%%%%%%%%%%%%%%%%%%%%%%%%%%%%%%%%%%%%%%%%%%%%

\begin{table}%[t]
  %\begin{center}
  \centering
    \begin{tabular}{|c|c||ccc|cc|cc|} \hline
      $z$ & noise & $E_0^{\delta \delta}$  & $E_2^{\delta \delta}$ & $E_4^{\delta \delta}$ & $E_1^{\delta u}$ & $E_3^{\delta u}$ & $E_0^{uu}$ & $E_2^{uu}$ \\ \hline\hline
      0.1 & \checkmark & 1.1 & 7.8 & 250 & 1.4 & 22 & 1.5 & 2.3\\
      0.1 & - & 1.0 & 7.5 & 240 & 1.4 & 21 & 1.3 & 2.2 \\
      0.5 & - & 1.0 & 5.9 & 140 & 1.4 & 17 & 1.3 & 2.2 \\
      1.0 & - & 1.0 & 5.2 & 110 & 1.4 & 15 & 1.3 & 2.2 \\
      2.0 & - & 1.0 & 4.9 & 94 & 1.4 & 14 & 1.3 & 2.2 \\
      3.0 & - & 1.0 & 4.8 & 90 & 1.4 & 13 & 1.3 & 2.2 \\
      \hline
    \end{tabular}
  %\end{center}
  \caption{Numerical values of the efficiency index $E_\ell^{X_1X_2}$ defined as the ratio of SNR for the TripoSH coefficient to that for the corresponding Legendre coefficient. We show seven possible $E_\ell^{X_1 X_2}$ at $b = 2$ in a LSST-level survey at $z = 0.1$ (first line) and a noiseless one at several redshift slices (second and subsequent lines).}\label{tab:SN}
\end{table}

\begin{figure}%[t]
%\begin{center}
  %% \includegraphics[width=0.47\textwidth]{SN_diag_WAbyLPP_per_z_b_nodip_CV_kmin1.0e-2hbyMpc_kmax1.0e-1hbyMpc.pdf}
    \includegraphics[width=\columnwidth]{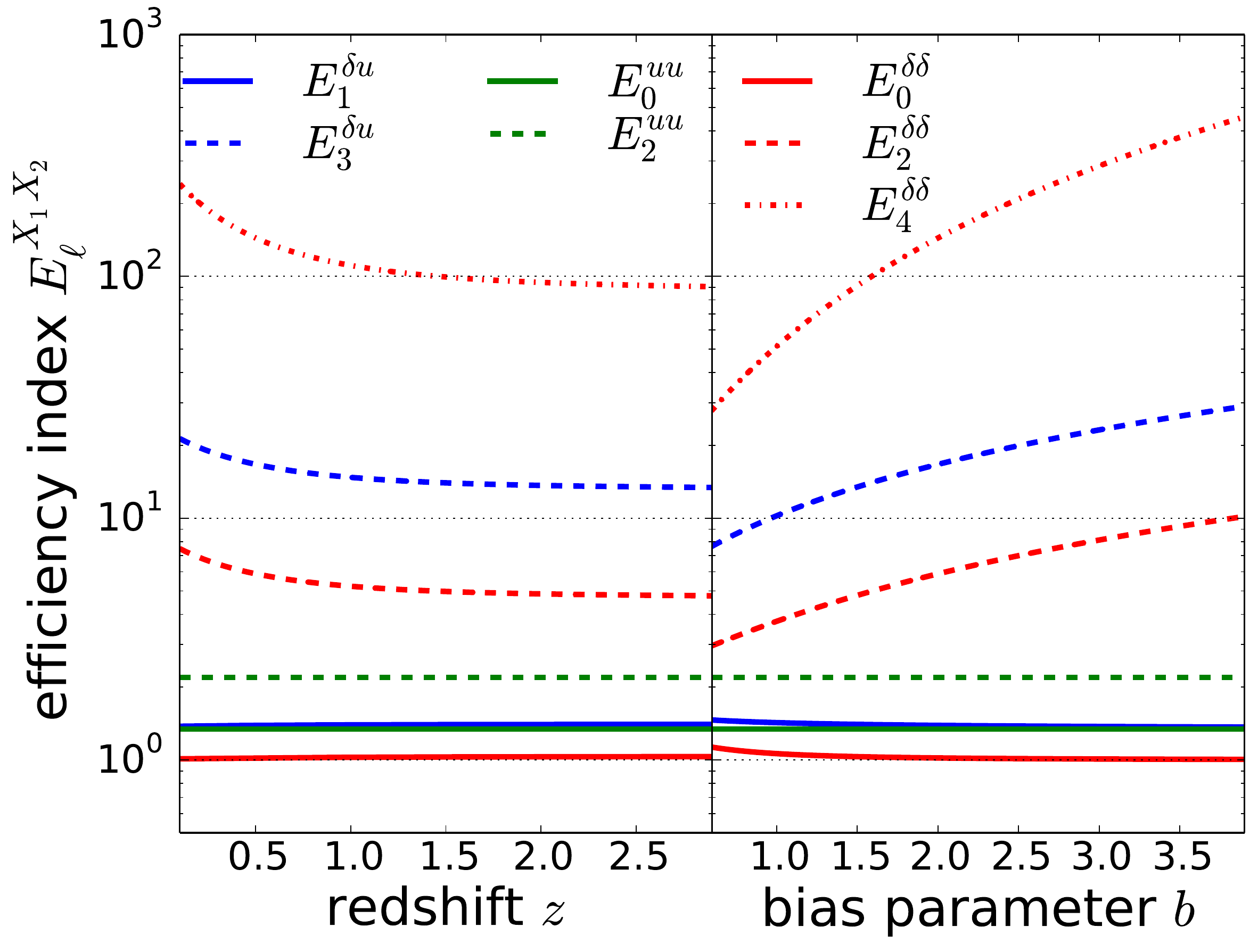}
%\end{center}
\caption{Seven possible $E_\ell^{X_1 X_2}$ in a noiseless survey as a function of $z$ at $b = 2$ (left) and a function of $b$ at $z = 0.5$ (right). }
\label{fig:SN}
\end{figure}

How much is the TripoSH decomposition [Eq.~\eqref{eq:Xi_def}] efficient? To see it, we estimate the SNR for each TripoSH coefficient according to
\begin{eqnarray}
  \left( \frac{S}{N} \right)_{\substack{X_1 X_2 \\ \ell \ell_1 \ell_2}}^2
  = V \int_{k_{\rm min}}^{k_{\rm max}} \frac{k^2 dk}{ 2\pi^2} 
  \frac{\left|{\cal P}_{\ell \ell_1 \ell_2}^{X_1 X_2}(k)\right|^2}{\Upsilon_{\ell \ell_1\ell_2 ; \ell \ell_1 \ell_2}^{X_1 X_2 ; X_1 X_2}(k)}.
\end{eqnarray}
In the possible $14+5+2$ coefficients listed in Eq.~\eqref{eq:calP_list}, the following $9+4+2$ ones form the minimum covariance \eqref{eq:Upsilon_CVL}:
\begin{eqnarray}
  \begin{cases}
    {\cal P}_{011}^{\delta \delta} , \  
    {\cal P}_{022}^{\delta \delta} , \\
    %@@@
    {\cal P}_{112}^{\delta \delta} , \
    {\cal P}_{121}^{\delta \delta} , \\ 
    %@@@
    {\cal P}_{211}^{\delta \delta} , \  
    {\cal P}_{222}^{\delta \delta} , \\
    %@@@
    {\cal P}_{312}^{\delta \delta} , \ 
    {\cal P}_{321}^{\delta \delta} , \\ 
    %@@@
    {\cal P}_{422}^{\delta \delta} ,
  \end{cases} 
    %---
  \begin{cases}
    {\cal P}_{011}^{\delta u} , \\
    %@@@
    {\cal P}_{121}^{\delta u} , \\
    %@@@
    {\cal P}_{211}^{\delta u} , \\
    %@@@
    {\cal P}_{321}^{\delta u} ,
  \end{cases}
    %---
  \begin{cases} 
    {\cal P}_{011}^{u u} , \\
    %@@@
    {\cal P}_{211}^{u u} .
  \end{cases}
  \label{eq:calP_noiseless_list}
\end{eqnarray}
Their SNRs are therefore maximally enhanced and take the identical form: 
\begin{eqnarray}
  \left( \frac{S}{N} \right)_{\substack{X_1 X_2 \\ \ell \ell_1 \ell_2}}^2
  = \frac{V}{ 12\pi^2}(k_{\rm max}^3 - k_{\rm min}^3 )  . \label{eq:SN2_CVL}
\end{eqnarray}
One can compute the SNR integrated over $\ell_1$ and $\ell_2$, while there is no additional gain and it reaches a ceiling of Eq.~\eqref{eq:SN2_CVL}.
 
Let us define the square root of Eq.~\eqref{eq:SN2_CVL} divided by the SNR of the corresponding Legendre coefficient as the efficiency index $E_\ell^{X_1 X_2}$, reading
  \begin{eqnarray}
    E_\ell^{X_1 X_2} \equiv \sqrt{ \frac{k_{\rm max}^3 - k_{\rm min}^3 }{6 \int_{k_{\rm min}}^{k_{\rm max}} k^2 dk \left|P_\ell^{X_1 X_2}(k)\right|^2 / \Theta_{\ell}^{X_1 X_2}(k)} } , 
  \end{eqnarray}
  where $\Theta_{\ell}^{X_1 X_2}(k)$ denotes the diagonal components of the covariance matrix of $\hat{P}_{\ell}^{X_1 X_2}$ as
\begin{eqnarray}
  \begin{split}
     \Theta_{\ell}^{X_1 X_2}(k)
  &\equiv N_k \Braket{\left| \hat{P}_{\ell}^{X_1 X_2}(k) \right|^2  }_c \\
  & =   
 \frac{(2\ell + 1)^2}{2} \int_{-1}^1 d(\hat{k} \cdot \hat{s})  \left[ {\cal L}_{\ell}(\hat{k} \cdot \hat{s}) \right]^2 \\ 
 & \quad \times \left[
   P_{{\rm tot}}^{X_1 X_1}({\bf k}, \hat{s}, \hat{s}) P_{{\rm tot}}^{X_2 X_2}(-{\bf k}, \hat{s}, \hat{s}) \right.  \\ 
   &\left. \qquad
   +  (-1)^{\ell}
    P_{\rm tot}^{X_1 X_2}({\bf k}, \hat{s}, \hat{s})
    P_{\rm tot}^{X_2 X_1}(-{\bf k}, \hat{s}, \hat{s})
    \right] .
  \end{split}
\end{eqnarray}
We compute seven possible indices: $E_0^{\delta  \delta}$, $E_2^{\delta  \delta}$, $E_4^{\delta  \delta}$, $E_1^{\delta u}$, $E_3^{\delta u}$, $E_0^{uu}$ and $E_2^{uu}$ at $k_{\rm min} = 0.01 h ~ {\rm Mpc^{-1}}$ and $k_{\rm max} = 0.1 h ~ {\rm Mpc^{-1}}$, by varying redshift $z$, the bias parameter $b$ and the noise power spectrum $P_{\rm noise}^{X_1 X_2}$. The values of the cosmological parameters adopted here are fixed to be consistent with the latest ${\it Planck}$ constraints \citep{Aghanim:2018eyx}. Note that $E_\ell^{X_1 X_2}$ is independent of $V$.

Some representative values at $b = 2$ are summarized in Table~\ref{tab:SN}. Here, a realistic survey at $z = 0.1$ (first line) and an ideal noiseless ones at several redshift slices (second and subsequent lines) are assumed. For the first case, we model the noise spectrum as $P_{\rm noise}^{\delta \delta} = 1/\bar{n}$, $P_{\rm noise}^{uu} = \sigma_u^2 /\bar{n}$ and $P_{\rm noise}^{\delta u} = P_{\rm noise}^{u \delta} = 0$, and assume $\sigma_u = 300~{\rm km/s}$ and $\bar{n} = 5 \times 10^{-4} h^3 ~ {\rm Mpc^{-3}}$ in anticipation of a LSST-level survey \citep{Graziani:2020kkr}. In contrast, for the second and subsequent cases, $P_{\rm noise}^{\delta \delta} = P_{\rm noise}^{uu} =  P_{\rm noise}^{\delta u} = P_{\rm noise}^{u \delta} = 0$ are adopted. One can see from this that $E_\ell^{X_1 X_2}$ is drastically enhanced at higher $\ell$. This is because, as explained above, the big contamination due to lower $\ell$ modes in the higher $\ell$ elements of the covariance matrix, which appears in the Legendre decomposition case, is completely absent in the TripoSH one. 

Figure~\ref{fig:SN} draws the dependence of $E_\ell^{X_1 X_2}$ on $z$ (left panel) and $b$ (right panel) assuming a noiseless survey. It is visually apparent that $E_2^{\delta  \delta}$, $E_4^{\delta  \delta}$ and $E_3^{\delta u}$ increase as $z$ decreases or $b$ increases. Decreasing $z$, namely decreasing the growth rate $f$, or increasing $b$ relatively enhances $P_0^{\delta \delta}$ and $P_1^{\delta u}$ as compared with $P_2^{\delta  \delta}$, $P_4^{\delta  \delta}$ and $P_3^{\delta u}$ and the contamination of the Legendre coefficient covariance gets worse, degrading the Legendre SNRs, i.e., the denominators of the efficiency indices. In contrast, the numerators, i.e., Eq.~\eqref{eq:SN2_CVL}, remain unchanged, and consequently $E_2^{\delta  \delta}$, $E_4^{\delta  \delta}$ and $E_3^{\delta u}$ are enhanced.  

The similar comparison can also be done in configuration space by computing the SNR of $\Xi_{\ell\ell_1\ell_2}^{X_1 X_2}$. Since $\Xi_{\ell\ell_1\ell_2}^{X_1 X_2}$ is linked with ${\cal P}_{\ell\ell_1\ell_2}^{X_1 X_2}$ through the simple Hankel transformation formula \eqref{eq:hankel}, the similar results are expected.

%%%%%%%%%%%%%%%%%%%%%%%%%%%%%%%%%%%%%%%%%%%%%%%%%%%%%%%%%%%%%%%%%%%%%%%%%%%%%%
\section{Summary and discussions}\label{sec:con}
%%%%%%%%%%%%%%%%%%%%%%%%%%%%%%%%%%%%%%%%%%%%%%%%%%%%%%%%%%%%%%%%%%%%%%%%%%%%%%

In this Letter, we have computed the covariance of the TripoSH-decomposed density/velocity power spectrum for the first time. We have shown that, by virtue of the complete angular decomposition using the TripoSH basis, nontrivial mode mixings at the covariance level, as seen in the usual Legendre decomposition, can be fully disentangled, and as a result, the covariance at each multipole mode is minimized. Via the simple SNR estimation, we have found that the detectability improvements by our decomposition approach are more significant for higher multipole modes, and there are some order of magnitude improvements at the hexadecapole of the density auto power spectrum and the octopole of the density-velocity cross one. In addition, odd (even) multipoles of the density auto (density-velocity cross) power spectrum, which vanish in the Legendre decomposition, are distinctive modes of the TripoSH one, producing additional gains of the total SNR. The obtained results encourage reanalyzing the currently available data based on the TripoSH decomposition approach instead of the Legendre one. Moreover, its application to the upcoming wide-angle galaxy surveys such as SPHEREx \citep{Dore:2014cca}, Euclid \citep{Laureijs:2011gra} and WFIRST \citep{Spergel:2013tha} is expected to further improve the detectability.

In the real data analysis, however, extra artificial signals due to specific survey geometry would remain as a residue to some extent even after the subtraction process and might make a nontrivial impact. Even in the theoretical analysis, the applicability of our covariance formula to the nonlinear regime \citep{Castorina:2018nlb,Taruya:2019xsf}, the general relativistic effect \citep[e.g.,][]{Bertacca:2012tp}, or the statistically-anisotropic Universe \citep{Shiraishi:2016wec} is not trivial.
Non-Gaussian contributions in the covariance, ignored in this Letter, could be brute-forcely estimated using mock galaxy catalogs. However, if the data vector is composed of all TripoSH coefficients, in comparison with the Legendre decomposition case \citep[according to][$\sim 10^3$]{White:2013psd,Kitaura:2015uqa}, a larger number of mocks will be required. If $\sim 10^4$ is enough, the brute-force computation will be still feasible. Otherwise, the extension of recently developed analytic approaches \citep[e.g.,][]{Wadekar:2019rdu} might resolve the difficulty. Together with the development of an efficient and feasible estimator of the TriposH coefficient and forecasting cosmological parameter constraints, these interesting issues will also be addressed in our future publications.

%%%%%%%%%%%%%%%%%%%%%%%%%%%%%%%%%%%%%%%%%%%%%%%%%%

\section*{Acknowledgements}

We thank Atsushi Taruya for useful discussion. M.S. and N.S.S are supported by JSPS Grant-in-Aid for Early-Career Scientists Grant Nos.~19K14718 and 19K14703, respectively. T.O. acknowledges support from the Ministry of Science and Technology of Taiwan under Grant Nos. MOST 106-2119-M-001-031-MY3 and MOST 109-2112-M-001-027- and the Career Development Award, Academia Sinica (AS-CDA-108-M02) for the period of 2019 to 2023. K.A. is supported by Grand-in-Aid for JSPS fellows No.~19J12254. M.S., N.S.S, and K.A. also acknowledge the Center for Computational Astrophysics, National Astronomical Observatory of Japan, for providing the computing resources of Cray XC50.

%%%%%%%%%%%%%%%%%%%%%%%%%%%%%%%%%%%%%%%%%%%%%%%%%%

%%%%%%%%%%%%%%%%%%%% Data Availability Statements %%%%%%%%%%%%%%%%%%

%% \section*{Data Availability Statements}

%% The data underlying this article are available in the article.

%%%%%%%%%%%%%%%%%%%%%%%%%%%%%%%%%%%%%%%%%%%%%%%%%%



\bibliographystyle{aasjournal}
\bibliography{paper} % if your bibtex file is called example.bib

\begin{thebibliography}{}
\expandafter\ifx\csname natexlab\endcsname\relax\def\natexlab#1{#1}\fi
\providecommand{\url}[1]{\href{#1}{#1}}
\providecommand{\dodoi}[1]{doi:~\href{http://doi.org/#1}{\nolinkurl{#1}}}
\providecommand{\doeprint}[1]{\href{http://ascl.net/#1}{\nolinkurl{http://ascl.net/#1}}}
\providecommand{\doarXiv}[1]{\href{https://arxiv.org/abs/#1}{\nolinkurl{https://arxiv.org/abs/#1}}}

\bibitem[{Aghanim {et~al.}(2018)}]{Aghanim:2018eyx}
Aghanim, N., {et~al.} 2018.
\newblock \doarXiv{1807.06209}

\bibitem[{Bertacca {et~al.}(2012)Bertacca, Maartens, Raccanelli, \&
  Clarkson}]{Bertacca:2012tp}
Bertacca, D., Maartens, R., Raccanelli, A., \& Clarkson, C. 2012, JCAP, 10,
  025, \dodoi{10.1088/1475-7516/2012/10/025}

\bibitem[{Beutler {et~al.}(2019)Beutler, Castorina, \& Zhang}]{Beutler:2018vpe}
Beutler, F., Castorina, E., \& Zhang, P. 2019, JCAP, 03, 040,
  \dodoi{10.1088/1475-7516/2019/03/040}

\bibitem[{Castorina \& White(2018{\natexlab{a}})}]{Castorina:2017inr}
Castorina, E., \& White, M. 2018{\natexlab{a}}, Mon. Not. Roy. Astron. Soc.,
  476, 4403, \dodoi{10.1093/mnras/sty410}

\bibitem[{Castorina \& White(2018{\natexlab{b}})}]{Castorina:2018nlb}
---. 2018{\natexlab{b}}, Mon. Not. Roy. Astron. Soc., 479, 741,
  \dodoi{10.1093/mnras/sty1437}

\bibitem[{Castorina \& White(2019)}]{Castorina:2019hyr}
---. 2019.
\newblock \doarXiv{1911.08353}

\bibitem[{Dor\'e {et~al.}(2014)}]{Dore:2014cca}
Dor\'e, O., {et~al.} 2014.
\newblock \doarXiv{1412.4872}

\bibitem[{Graziani {et~al.}(2020)}]{Graziani:2020kkr}
Graziani, R., {et~al.} 2020.
\newblock \doarXiv{2001.09095}

\bibitem[{Hamilton(1997)}]{Hamilton:1997zq}
Hamilton, A. 1997, in {Ringberg Workshop on Large Scale Structure},
  \dodoi{10.1007/978-94-011-4960-0\_17}

\bibitem[{Kitaura {et~al.}(2016)}]{Kitaura:2015uqa}
Kitaura, F.-S., {et~al.} 2016, Mon. Not. Roy. Astron. Soc., 456, 4156,
  \dodoi{10.1093/mnras/stv2826}

\bibitem[{Laureijs {et~al.}(2011)}]{Laureijs:2011gra}
Laureijs, R., {et~al.} 2011.
\newblock \doarXiv{1110.3193}

\bibitem[{{Matsubara} {et~al.}(2000){Matsubara}, {Szalay}, \&
  {Landy}}]{Matsubara:2000a}
{Matsubara}, T., {Szalay}, A.~S., \& {Landy}, S.~D. 2000, \apjl, 535, L1,
  \dodoi{10.1086/312701}

\bibitem[{{Okumura} {et~al.}(2008){Okumura}, {Matsubara}, {Eisenstein}, {Kayo},
  {Hikage}, {Szalay}, \& {Schneider}}]{Okumura:2008}
{Okumura}, T., {Matsubara}, T., {Eisenstein}, D.~J., {et~al.} 2008, \apj, 676,
  889, \dodoi{10.1086/528951}

\bibitem[{Papai \& Szapudi(2008)}]{Papai:2008bd}
Papai, P., \& Szapudi, I. 2008, Mon. Not. Roy. Astron. Soc., 389, 292,
  \dodoi{10.1111/j.1365-2966.2008.13572.x}

\bibitem[{{Peebles}(1980)}]{Peebles:1980}
{Peebles}, P.~J.~E. 1980, {The large-scale structure of the universe}
  (Princeton, N.J., Princeton Univ. Press)

\bibitem[{{Pope} {et~al.}(2004){Pope}, {Matsubara}, {Szalay}, {Blanton},
  {Eisenstein}, {Gray}, {Jain}, {Bahcall}, {Brinkmann}, {Budavari}, {Connolly},
  {Frieman}, {Gunn}, {Johnston}, {Kent}, {Lupton}, {Meiksin}, {Nichol},
  {Schneider}, {Scranton}, {Strauss}, {Szapudi}, {Tegmark}, {Vogeley},
  {Weinberg}, {Zehavi}, \& {SDSS Collaboration}}]{Pope:2004}
{Pope}, A.~C., {Matsubara}, T., {Szalay}, A.~S., {et~al.} 2004, \apj, 607, 655,
  \dodoi{10.1086/383533}

\bibitem[{Shiraishi {et~al.}(2017)Shiraishi, Sugiyama, \&
  Okumura}]{Shiraishi:2016wec}
Shiraishi, M., Sugiyama, N.~S., \& Okumura, T. 2017, Phys. Rev. D, 95, 063508,
  \dodoi{10.1103/PhysRevD.95.063508}

\bibitem[{Spergel {et~al.}(2013)}]{Spergel:2013tha}
Spergel, D., {et~al.} 2013.
\newblock \doarXiv{1305.5422}

\bibitem[{{Strauss} \& {Willick}(1995)}]{Strauss:1995}
{Strauss}, M.~A., \& {Willick}, J.~A. 1995, \physrep, 261, 271,
  \dodoi{10.1016/0370-1573(95)00013-7}

\bibitem[{Szalay {et~al.}(1998)Szalay, Matsubara, \& Landy}]{Szalay:1997cc}
Szalay, A.~S., Matsubara, T., \& Landy, S.~D. 1998, Astrophys. J., 498, L1,
  \dodoi{10.1086/311293}

\bibitem[{Szapudi(2004)}]{Szapudi:2004gh}
Szapudi, I. 2004, Astrophys. J., 614, 51, \dodoi{10.1086/423168}

\bibitem[{Taruya {et~al.}(2010)Taruya, Nishimichi, \& Saito}]{Taruya:2010mx}
Taruya, A., Nishimichi, T., \& Saito, S. 2010, Phys. Rev. D, 82, 063522,
  \dodoi{10.1103/PhysRevD.82.063522}

\bibitem[{Taruya {et~al.}(2020)Taruya, Saga, Breton, Rasera, \&
  Fujita}]{Taruya:2019xsf}
Taruya, A., Saga, S., Breton, M.-A., Rasera, Y., \& Fujita, T. 2020, Mon. Not.
  Roy. Astron. Soc., 491, 4162, \dodoi{10.1093/mnras/stz3272}

\bibitem[{Varshalovich {et~al.}(1988)Varshalovich, Moskalev, \&
  Khersonsky}]{Varshalovich:1988ye}
Varshalovich, D., Moskalev, A., \& Khersonsky, V. 1988, {Quantum Theory of
  Angular Momentum: Irreducible Tensors, Spherical Harmonics, Vector Coupling
  Coefficients, 3nj Symbols} (Singapore: World Scientific)

\bibitem[{Wadekar \& Scoccimarro(2019)}]{Wadekar:2019rdu}
Wadekar, D., \& Scoccimarro, R. 2019.
\newblock \doarXiv{1910.02914}

\bibitem[{Weinberg {et~al.}(2013)Weinberg, Mortonson, Eisenstein, Hirata,
  Riess, \& Rozo}]{Weinberg:2012es}
Weinberg, D.~H., Mortonson, M.~J., Eisenstein, D.~J., {et~al.} 2013, Phys.
  Rept., 530, 87, \dodoi{10.1016/j.physrep.2013.05.001}

\bibitem[{White {et~al.}(2014)White, Tinker, \& McBride}]{White:2013psd}
White, M., Tinker, J.~L., \& McBride, C.~K. 2014, Mon. Not. Roy. Astron. Soc.,
  437, 2594, \dodoi{10.1093/mnras/stt2071}

\bibitem[{Yamamoto {et~al.}(2006)Yamamoto, Nakamichi, Kamino, Bassett, \&
  Nishioka}]{Yamamoto:2005dz}
Yamamoto, K., Nakamichi, M., Kamino, A., Bassett, B.~A., \& Nishioka, H. 2006,
  Publ. Astron. Soc. Jap., 58, 93, \dodoi{10.1093/pasj/58.1.93}

\bibitem[{Yoo \& Seljak(2015)}]{Yoo:2013zga}
Yoo, J., \& Seljak, U.~s. 2015, Mon. Not. Roy. Astron. Soc., 447, 1789,
  \dodoi{10.1093/mnras/stu2491}

\end{thebibliography}

%%%%%%%%%%%%%%%%%%%%%%%%%%%%%%%%%%%%%%%%%%%%%%%%%%

%%%%%%%%%%%%%%%%% APPENDICES %%%%%%%%%%%%%%%%%%%%%


%% \section{Some extra material}

%%%%%%%%%%%%%%%%%%%%%%%%%%%%%%%%%%%%%%%%%%%%%%%%%%

% Don't change these lines
\bsp	% typesetting comment
\label{lastpage}
\end{document}